\documentclass[a4paper,12pt,oneside,onecolumn]{article}
\usepackage{color,graphicx}

\usepackage{geometry}%
\definecolor{c1}{rgb}{1,0,0}
\definecolor{c2}{rgb}{0,0,1}
\usepackage{titlesec}%

\newcommand{\ppd}[2]{\frac {\partial #1 }{\partial #2}}

\newcommand{\pdd}[2]{\frac {\partial^2 #1 }{\partial {#2}^2}}
\usepackage[colorlinks=true,linkcolor=black, 
filecolor=black, 
citecolor=black,
urlcolor=blue]{hyperref}

\newcommand{\cor}[1]{#1}%

\title{A simple and efficient model for mesoscale solidification simulation of globular grain structures}
\author{St\'ephane Vern\`ede$^{1,2}$, Michel Rappaz$^1$}%
\date{12 october 2006}
\begin{document}

\maketitle
\begin{center}
\begin{small}
 
$^1$ Computational Materials Laboratory,\\
Ecole Polytechnique F\'ed\'erale de Lausanne,\\ Station 12, Lausanne, CH-1015 Switzerland\vspace{5 mm }\\
$^2$Alcan-CRV,\\ ZI Centr'alp, 0725 rue Aristide Berges,\\ BP 27, Voreppe, FR-38341 France
\end{small}
\end{center}
\begin{center}
keywords:  solidification, equiaxed microstructure, modelling, surface energy\\
\end{center}
\begin{center}
Published in Acta Materialia, Volume 55, Issue 5, March 2007, Pages 1703-1710\\
\href{http://dx.doi.org/10.1016/j.actamat.2006.10.031}{view at publisher}

\end{center}
\begin{abstract}

A simple model for the solidification of globular grains in
metallic alloys is presented. Based on the Voronoi diagram of the
nuclei centers, it accounts for the curvature of the grains near
triple junctions. The predictions of this model are close to those
of more refined approaches such as the phase field method, but
with a computation cost decreased by several orders of magnitude.
Therefore, this model is ideally suited for \textit{granular
simulations} linking the behavior of individual grains to
macroscopic properties of the material.

\end{abstract}

\section{Introduction}
Microstructures that form during solidification of metallic alloys
play a key role for the final properties of as-cast materials and
for subsequent heat treatments. They also condition the formation
of defects such as porosity and hot cracking \cite{Campbell1991}. As
solidification of one or more solid phases is a moving
free-boundary problem, several techniques have been developed
especially over the past decade to overcome the difficulty of
tracking the solid-liquid interface.

One of the most powerful and widely used methods is the phase
field technique \cite{phase_field_rev}. In this method, the sharp
interface is replaced by a continuous field varying from 0 in the
solid to 1 in the liquid with a diffuse interface over a finite
thickness. As the phase field method is based on thermodynamics
considerations, it provides a unified framework for many phenomena
(formation of dendrites, eutectics, peritectics, etc.). Yet, this
method is very computation intensive as the mesh size has to be
small with respect to the diffuse interface thickness, and the
diffuse interface thickness has to be small with respect to the
typical radius of curvature of the microstructure (e.g., dendrite
tip radius, eutectic spacing). Moreover, since the problem is
generally solved using an explicit scheme, a Fourier criterion
must be satisfied to ensure numerical stability. This is why
nowadays, phase field simulations are limited to the simulation of
a few grains, very often in two dimensions. Other techniques such
as the level set \cite{jon_level_set} or pseudo-front tracking
\cite{jacot} methods have been developed, with different
advantages and drawbacks, but they roughly require the same amount
of computations.

If fundamental aspects of microstructure formation can be tackled
with such sophisticated methods, it must also be recognized that
there is an increasing need for simulations that can relate
microstructure and macroscopic properties. This is particularly
important for the last stage of solidification, during which the
most important defects (porosity, hot tears) form. In particular,
the gradual topological transition of the microstructure from
continuous liquid films to a continuous and fully coherent solid
is essential for hot tearing. Because of the random nature of
nucleation, this transition is associated with the formation of
increasingly larger clusters of grains as described by the
percolation theory \cite{percotheorie}. Therefore, the mechanical
properties and feeding ability of the mushy zone during this
transition cannot be deduced from the morphology of just a few
grains \cite{second}.

In order to study the gradual formation of a coherent solid phase
in globular alloys, an original approach has been proposed by
Mathier et al. \cite{vince2}. In this model, grains are
approximated by polyhedrons based on the Voronoi diagram of a
random set of nuclei. In order to compute the solidification of
large and non-isothermal mushy zones, Vern\`ede et al. further
simplified  the assumptions of this model  \cite{premier}. Yet,
these two approaches lead to polyhedral grains and cannot account
for the formation of isolated liquid pockets at the triple
junctions of the grains.

In the present contribution, a solidification model based on the
approach of Mathier et al. but accounting for the Gibbs-Thomson
effect near grain corners is described. The predictions of this
model are shown to be close to those obtained with more refined
methods such as the phase field or pseudo-front tracking methods,
but with a computation time several orders of magnitude lower.
After a brief recall of the \textit{granular model }of
solidification based on the Voronoi tessellation, the account of
the Gibbs-Thomson effect and its implementation in the model are
described. In the last section, simulation results are compared
with those obtained with a pseudo-front tracking method, thus
providing a basis of discussion for setting up bonds to this
simple approach.

\section{A solidification model based on Voronoi diagrams}

This 2 dimensional solidification model has been  derived first by
Mathier et al. \cite{vince2} and further developed by Vern\`ede et
al. \cite{premier}. \cor{ For a better understanding the main lines
 of this model are recalled here.}

\begin{figure}
\begin{center}
\includegraphics[height=6.2 in, angle=-90]{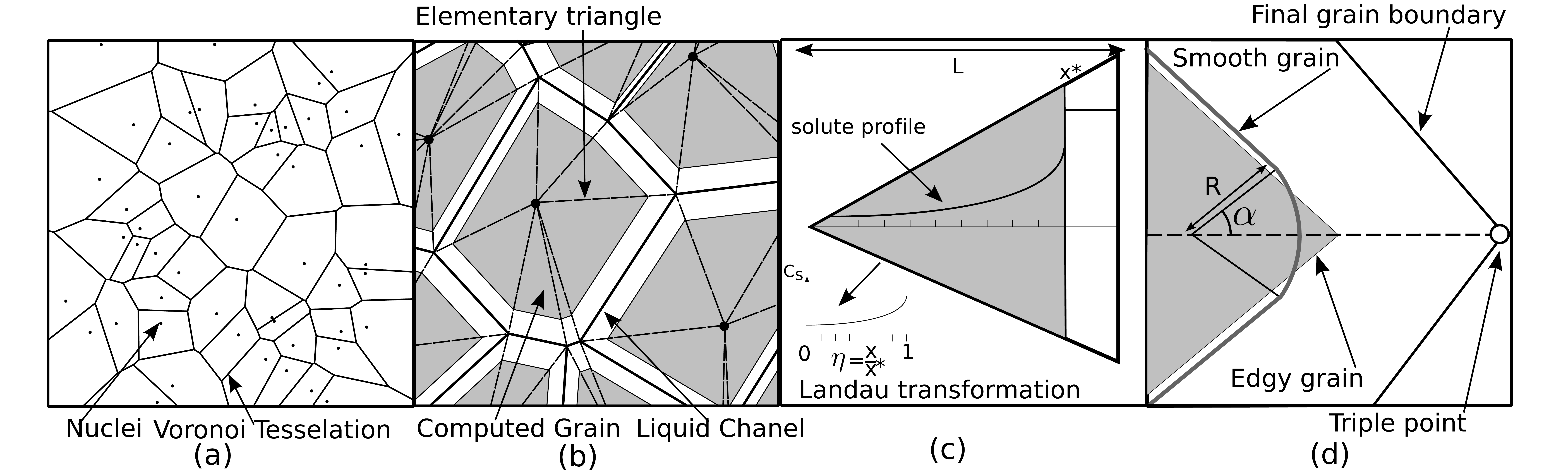}
\caption{Various enlargements of the granular model: Voronoi
tessellation associated with the nuclei centers (a); shape of the
grains during solidification (b); solute balance within one
triangle (c); smoothing procedure of the solid-liquid interface
near the grain corners (d).} \label{solidif}
\end{center}
\end{figure}

The model assumes simultaneous nucleation of grains in a plane
with a given density of random sites. Further assuming that the
temperature difference across the average grain size is small with
respect to the undercooling (i.e., small thermal gradient), the
final grain structure is close to the Voronoi tessellation of the
set of nuclei (Fig. \ref{solidif}, (a)) \cite{charbon}. In the
present work, the Voronoi tessellation was computed using the free
access software \textit{qhull} \cite{qhull}.

In order to further simplify the solidification model, the solute
flux between elementary triangles is neglected in a first step.
Thus, the smooth interface of each grain during growth can be
approximated by a linear segment in each triangle connecting the
nucleation center with a Voronoi segment. By construction, these
segments are perpendicular to the vectors connecting the
nucleation centers and  the two triangles issued from the same
Voronoi segment are symmetrical, (Fig. \ref{solidif}, (b)).

Solidification is reduced therefore to a one-dimensional problem
in each triangle, with the assumption of complete mixing of solute
in the liquid phase and back-diffusion in the solid. Moreover, the
temperature of the system is imposed, either uniform or given by a
fixed thermal gradient and decreasing with a given cooling rate.
\cor{Therefore, the solute balance integrated over the liquid phase of an elementary triangle gives:}
\begin{equation}
x^* D_s {\partial c_s^* \over
\partial x}(x^*)+ v^*x^*(k-1)c_\ell+\frac{1}{2}(L^2-x^{*^2})\ppd{c_\ell}{t}=0
\label{s_balance}
\end{equation}
where $x^*$ and $v^*$ are the position and speed of the interface,
respectively, $c_s$ and $c_\ell$ the solute concentration in the
solid and liquid phase, $c_s^*$ the solute concentration in the
solid at the interface, $k$ the partition coefficient, $t$ the time, $D_s$ the diffusion coefficient in the solid,
 $L$ the height of the elementary triangle perpendicular to the Voronoi
segment and $x$ the coordinate along this direction
(Fig. \ref{solidif}, (c)).\cor{ The last term of Eq. \ref{s_balance} accounts for the evolution of solute 
concentration, which is imposed by an external cooling rate $\dot{T}$ and the phase diagram. Thus}  
\begin{equation}
x^* D_s {\partial c_s^* \over
\partial x}(x^*)+ v^*x^*(k-1)c_\ell+\frac{1}{2}(L^2-x^{*^2})\frac{\dot{T}}{m}=0
\label{interf}
\end{equation}
where $m$ is  the slope of the liquidus.
 
The first term in Eq. \ref{interf} associated with back-diffusion in the solid is computed by solving
the cylindrical diffusion equation in the solid phase, \cor{which can be viewed as the solute balance between slices of the elementary triangle for constant $x$: 
\begin{equation}
\ppd{c_s}{t} = D_s \left(  \pdd{c_s}{x}+ \frac{1}{x} \ppd{c_s}{x}  \right) 
\label{difu}
\end{equation}
}
\cor{In order to easily account for solidification, we use a  Landau transformation of the
solid domain [0,x*(t)] into the reference 1D domain [0,1], as introduced by Voller and Sundarraj \cite{Landau_solidif}:
\begin{equation}
c_s(x,t) \rightarrow c_s(\eta,t)~~~\eta=\frac{x}{x^*}
\end{equation}
Thus Eq. \ref{difu} becomes:
\begin{equation}
\left( \ppd{c_s}{t}\right) _{\eta} = \frac{D_s}{x^{*2}}   \pdd{c_s}{\eta}+ \left( \frac{\eta v^*}{x^*}+ \frac{D_s}{\eta x^{*2}} \right)  \ppd{c_s}{\eta}  
\end{equation}
where the term  $\frac{\eta v^*}{x^*}$ accounts for the advection of the mesh with the solidification front.
This equation is solved using a finite difference scheme, with a nil flux condition for $\eta=0$ and an imposed concentration $c_s^*=k c_l$ for $\eta=1$.}\cor{ Knowing the flux associated with back-diffusion and the concentration evolution in the liquid, the second term of Eq.  \ref{interf} allows to deduce the velocity of the interface, $v^*$, and thus to find the new position $x^*(t+dt)$.}

Note that solidification does not
depend on the opening of the elementary triangles but only on its
height $L$, i.e, on the half-distance between two nucleation
centers. Near the end of solidification, the excess free energy needed to
form a grain boundary out of two misoriented grains, i.e.,
coalescence undercooling \cite{coal}, can be accounted for. This
procedure is presented in Refs \cite{vince2,premier} \cor{, but as this feature is not useful
for the present contribution, it will not be detailed here.}

\section{A model for grain corners}
\label{model}

Although the previous model predicts fairly well the evolutions of
the grains, of the solid-liquid interface and of the solid
fraction (see section \ref{valid}), it leads to polyhedral grains
(Fig. \ref{solidif}, (d)). This situation is clearly unrealistic
as sharp corners are normally remelted by surface tension in
non-faceted crystals. Furthermore, no liquid pocket can form at
the triple junctions of the grains. As a consequence, this model
overestimates the volume fraction of solid at which contact
between grains occurs, as compared with experiments \cite{premier,
organiques}.

In order to remove these limitations, the Gibbs-Thomson effect at
grain corners has to be considered. Assuming again that
temperature is homogeneous at the scale of a grain, the liquid
concentration at an interface with a local radius of curvature $R$
is given by:

\begin{equation}
c_\ell^R=c_\ell^{\infty}+\frac{\Gamma_{s\ell}}{Rm} \label{gt}
\end{equation}
where $c_\ell^{\infty}$ is the liquid solute concentration for a
flat interface and $\Gamma_{s\ell}$ is the Gibbs-Thomson
coefficient. (Note that, in general, the liquidus slope $m$ is
negative). If solidification would be arrested, solute will flow
from low curvature areas to high curvature areas, as for
coarsening, thus remelting the highly curved zones. Yet, globular
grains are not spherical during solidification and another
phenomenon should balance the Gibbs-Thomson effect.

\begin{figure}[h]
\begin{center}
\includegraphics[height=3 in]{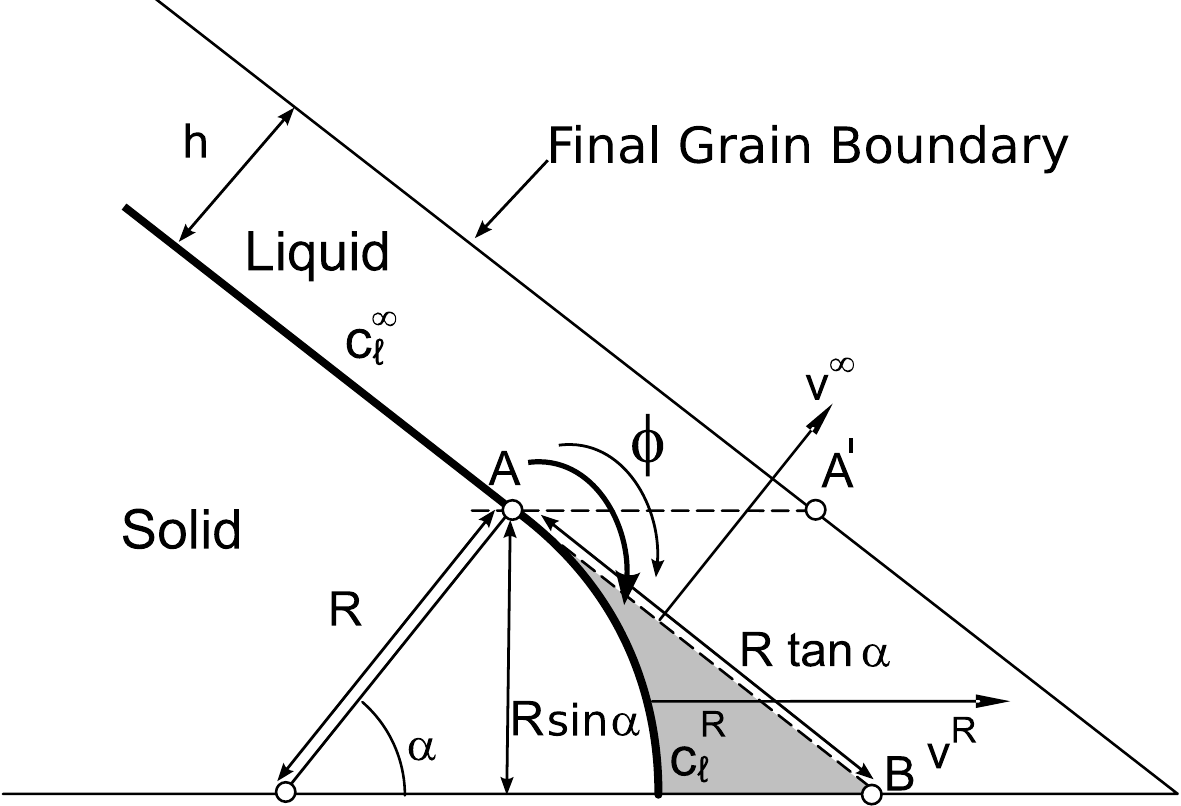}
\caption{Schematics of the solute flux model which accounts for
the Gibbs-Thomson effect near a grain corner.} \label{coin}
\end{center}
\end{figure}

Therefore, the idealized situation represented in Fig. \ref{coin}
is considered for smoothing the shape of polyhedral grains based
on the Gibbs-Thomson effect. The grain corner is modelled by a
curved interface with a constant radius of curvature $R$, whereas
elsewhere the interface is supposed flat and parallel to the final
grain boundary. Moreover, the liquid is divided into two zones,
delimited by the dotted line AA' in Fig. \ref{coin}. This line, passes through the point
separating the flat and circular portions of the solid-liquid
interface (point labelled A in Fig. \ref{coin}).
Close to the corner, the liquid has a concentration equal to $c_\ell^R$, whereas in the second zone,
surrounding the planar interface, a homogeneous concentration equal to
$c_\ell^\infty$ is considered.

The gradient of solute around the corner can be estimated by $(c_\ell^R - c_\ell^\infty)/(R \sin \alpha)$ 
where $R \sin \alpha$ is the distance between the tips of the corner and limit of the two zones. Moreover, we consider that the flux of solute is effective over a length of the order of $R \sin \alpha$. This leads to an estimation of the overall flux flowing from the zone surrounding the flat interface to the zone surrounding the corner:
\begin{equation}
\Phi \sim  D_\ell { (c_\ell^\infty- c_\ell^R) } = -{D_\ell
\Gamma_{s\ell} \over m R} > 0\label{flux}
\end{equation}
We further look for a radius of curvature that is stable with time , i.e., $\dot{R}=0$. This implies that the limit between the curved and planar interface stays along the dotted line AA' in Fig \ref{coin}. Once this constrain is fixed, it can be shown that the rejection of solute associated with a flat interface moving at a velocity $v^\infty$ is equal to that rejected by the projection of the curved interface moving at a velocity $v^R$. Indeed, the projection of the curved interface is given by $R \sin \alpha $, whereas the velocity is given by $v^R=v^\infty/ \cos \alpha$. So one has:
\begin{equation}
v^\infty R \tan \alpha = R \sin \alpha v^R = R \sin \alpha{v^\infty \over \cos \alpha}
\label{eq_flux}
\end{equation}
where $R \tan \alpha$ is the length of the extended flat interface
(dotted line AB in Fig. \ref{coin}). However, the grey surface is an additional volume
of liquid, compared to the flat interface, and represents the advantage
of the corner for solute diffusion. In other words, the small incoming 
flux of solute contributes to increasing the concentration of the liquid in the
area, $S$, of the grey zone. A more rigorous development, based on solute balances, 
is given in Appendix. But in summary, one has:
\begin{equation}
\Phi = {dc_\ell^R \over dt} S = {dc_\ell^\infty \over dt} S =
{\dot{T} \over m} S = {\dot{T} \over m} {R^2 \over 2} (\tan \alpha
- \alpha) \label{flux_balance}
\end{equation}
Combining Eq. \ref{flux} and \ref{flux_balance}, one finally gets:
\begin{equation}
R^3 = A_C {2 \over \tan \alpha - \alpha} {\Gamma_{s\ell} D_\ell
\over -\dot{T}} \label{R}
\end{equation}
where $A_C$ is a dimensionless constant, arising from the simple description 
of the solute distribution (see Eq. \ref{flux} ). Nonetheless, simply setting its value to 1  
produce satisfying results and this value will be used hereafter.
 
It is interesting to note at this stage that Eq. \ref{R} is close to a coarsening law:
the radius of curvature of a grain corner is proportional to the
third power of a driving force given by ${\Gamma_{s\ell} D_\ell
\over \Delta T_o}$, where $\Delta T_o$ is the solidification
interval of the alloy, and to the third power of the
solidification time, $t_f$. The geometrical factor in front of
this term is such that the radius of curvature becomes infinite  when $\alpha$ is equal to $0$ (the grain corner is
flat) and nil when $\alpha$ is equal to ${\pi \over 2}$ (the grain has disappeared).

Using this relationship for the radius of curvature of the grain
corners, the shape of the solid-liquid interface can be computed
as follows.
\begin{itemize}
\item The position of the flat interface or solid fraction is
computed for each elementary triangle using the back-diffusion
model described above \item The radius of curvature at each grain
corner is computed using Eq. \ref{R}. Please note that, for a
fixed cooling rate, the radius is constant and can be computed
only once before the time stepping procedure. \item The interface
in each elementary triangle is approximated by a flat portion and
two rounded corners. If the length of the flat interface becomes
negative, the interface is approximated by an arc of a circle.
\item Rounded interfaces increase the overall volume of liquid by
creating liquid pockets at grain corners. The position of the flat
interface is then moved slightly forward in order to conserve the
solid fraction computed with the flat interface method (see Fig.
\ref{solidif} (d)).
\end{itemize}

\section{Validity and limits of the model}
\label{valid}

The predictions of the present model have been compared with those
obtained with a pseudo front tracking method (PFT) \cite{jacot}.
In this technique, the fraction of solid within each cell of an
hexagonal network is computed based on an explicit solute
diffusion calculation. A layer of cells always separate the solid
and liquid phases and the solute flux balance for such cells is
converted into a solid fraction evolution. The position of the
interface within these interfacial cells is computed using a
piece-wise linear interface calculation (PLIC) algorithm \cite{plic}. Once the interface position within
each interfacial cell is known, its curvature and the associated
Gibss-Thomson effect are calculated using a distance-field method
similar to level-set. This method leads to predictions close to
the phase field method as shown in more details in Ref.
\cite{jacot}.

The same simulations have been carried out with the PFT and the
Voronoi methods. Six nuclei have been randomly placed in a $4.5\cdot 
10^{-2}~\mathrm{mm^2}$ domain with periodic boundary conditions.
The average grain size $d_c$ is therefore around 90 $\mu$m. An
Al-1wt$\%$Cu has been considered with a linearised phase diagram
($m = -6.67$ K/wt$\%$ and $k = 0.14$). The other parameters used
in these calculations are: $D_s = 1.5\cdot 10^{-13}$ ${\mathrm{ m^2/s}
}$, $D_\ell = 10^{-9}\cdot \mathrm{ m^2/s }$, $\Gamma_{s\ell} =
5\cdot 10^{-7}$ Km, $\dot{T}$ = -1 K/s.

\begin{figure}
\begin{center}
\includegraphics[height=2.5 in]{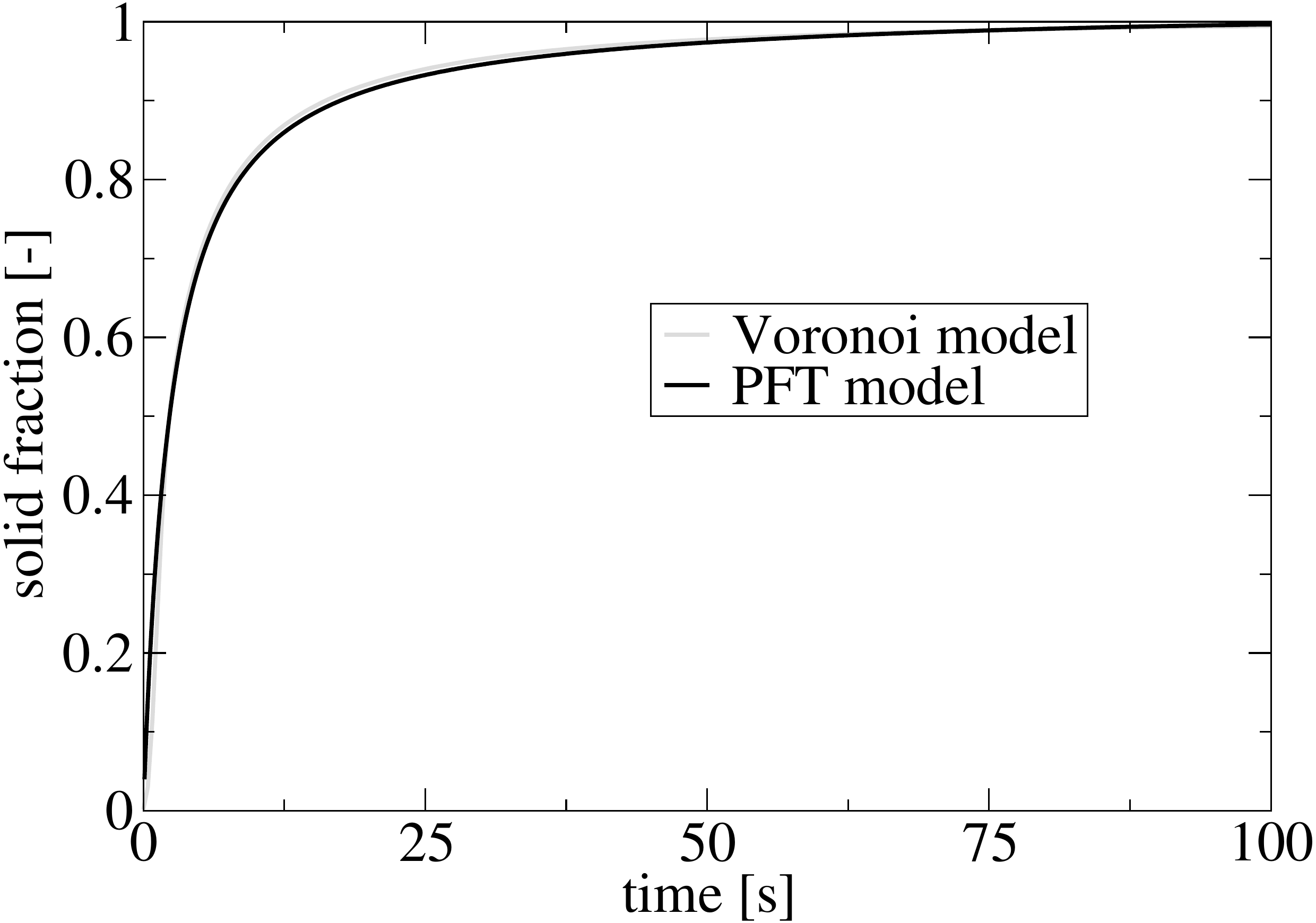}
\caption{Solid fraction as a function of time computed with the
Voronoi and the PFT methods} \label{fs}
\end{center}
\end{figure}

Figure \ref{fs} shows the evolution of the solid fraction, $g_s$,
computed with the two methods. Please keep in mind that, in the
Voronoi method, the solid fraction calculated with rounded grains
is equal to that obtained with polygonal grains.  At the very
beginning of solidification, grain growth predicted with the PFT
method is slightly faster as the Voronoi method assumes complete
mixing of solute in the liquid. As soon as a state of complete mixing is
reached with the PFT method, the predictions of the two models are
in very good agreement, even at high solid fraction. This shows
that, despite the fairly strong assumptions of the (flat
interface) Voronoi model, solute back-diffusion is well
approximated.

\begin{figure}
\begin{center}
\includegraphics[height=\textwidth, angle=-90]{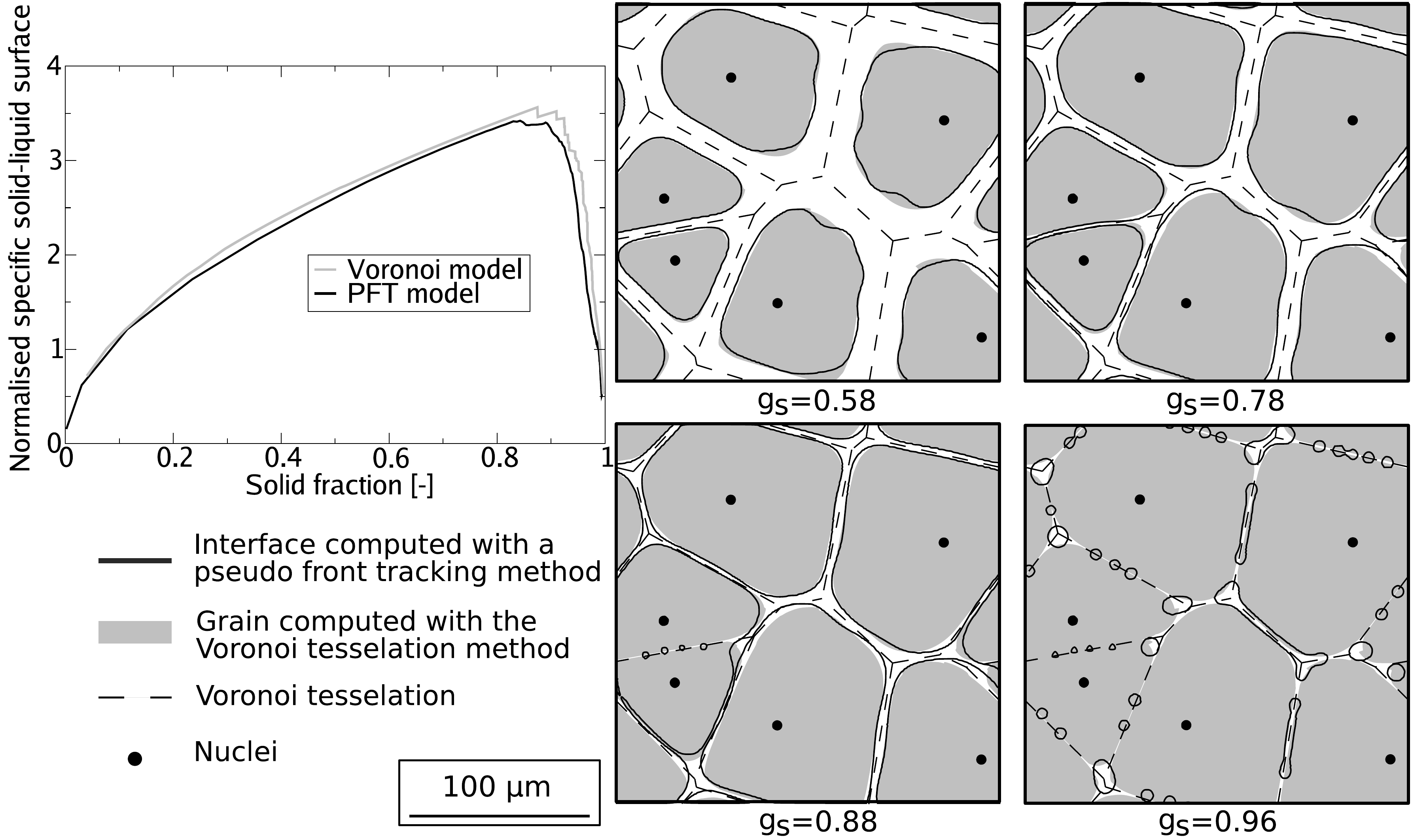}
\caption{Comparison of the solid liquid interface shape predicted
by the Voronoi and the PFT methods at various solid fractions.}
\label{interfaces}
\end{center}
\end{figure}

In order to compare in more details the shape of the interfaces
predicted by the two methods, the normalized specific solid liquid
interface, $S_s^{\circ}$, is represented in Fig \ref{interfaces}
as a function of $g_s$. This dimensionless number corresponds to
the total length of the solid-liquid interface of the grains
divided by the number of grains and by the average final grain
size $d_c$. This important parameter strongly influences the
permeability of the grain assembly via the Carman-Kozeny
relationship \cite{Carman} as verified experimentally in aluminium
alloys by Nielsen et al. \cite{Ksv}. A numerical study of
$S_s^{\circ}$ also shows that this parameter is an indicator of
the morphological transitions of the mushy zone \cite{second}. As
solid grains grow, the length of the solid liquid interface
increases as $\sqrt{g_s}$ in 2D until impingement/contact of the
solid grains makes it go to zero. Thus, $S_s^{\circ}$ is maximum
when the new contacts between the grains counterbalance the
natural increase of the interface length.

As can be seen on the left of Fig. \ref{interfaces}, the overall
shapes of $S_s^{\circ}(g_s)$ are in excellent agreement, but the
Voronoi method slightly overestimates this parameter, thus
revealing that grains are still slightly less rounded than those
calculated with the PFT method. It should be specified that, under
these conditions, the grains are clearly globular without
significant destabilization of the interface. Formation of
dendritic or globular dendritic grains would definitely increase
the length of the solid-liquid interface, thus making
$S_{s,PFT}^{\circ} > S_{s,Vor}^{\circ}$.

The grain shapes predicted by the two models at various solid
fractions are also represented in Fig. \ref{interfaces}. For
visualization purpose, the grains computed by the Voronoi method
are represented in grey, whereas the interfaces predicted by the
PFT method are represented with black lines. Again, a fairly good
agreement between the two simulations, especially near the grain
corners, can be seen. Please note that interfaces predicted with
the Voronoi model are not necessarily continuous as solidification
is computed separately for each elementary triangle. Yet, these
slight discontinuities do not affect much the topology of the
liquid channel network, nor the estimation of the channels
permeability and mushy zone topology.

The maximum of $S_s^{\circ}$ is predicted to occur at $g_s = 0.86$
by the PFT model, while it is delayed to $g_s = 0.89$ with the
Voronoi model. This difference can be easily understood by looking
at the shape of the grains at $g_s=0.88$. With the PFT model, the
liquid film in between two close neighboring grains can break down
 into small droplets by coalescence, whereas with the Voronoi
model it remains a film until final impingement. Such a liquid
film instability occurs when the grain boundary energy is lower
than twice the solid-liquid interfacial energy, i.e., attractive
boundaries, and the film thickness is on the order of the
thickness of the diffuse solid-liquid interfaces.\cite{coal} No
grain boundary energy has been set up in the PFT and Voronoi
calculations, but the instability occurs too early with the PFT
method as it is conditioned by the mesh size rather than the
diffuse interface thickness. Note that the quantitative simulation 
of the coalescence phenomenon remains a challenging topic despite 
advances done in the thermodynamic 
framework  \cite{coal} and in the modelling of triple phase boundaries \cite{triple_point}. Please note also that grain boundary
energies, and thus coalescence undercooling of repulsive
interfaces, can be easily introduced in the present
simulations.\cite{premier}

Despite the numerical difficulties associated with simulation of
the last stage solidification, the radii of curvature at grain
corners calculated with Eq. \ref{R} are close to those arising
from the complex PFT calculation, even for relatively narrow
channels. After breakdown of the liquid films, the sizes of the
liquid pockets at triple junction points are also in fairly good
agreement. Note that the isolated liquid pocket are represented
nevertheless with a negative curvature in the Voronoi simulation.
It is clear that the rounding procedure does not account for the
coalescence phenomenon which transforms globally convex globular
grains into convex liquid pockets. This simplified approach
nevertheless allows to introduce realistic volumes of liquid at
triple junctions, thus decreasing the volume fraction of solid at
which the flat portions of the grain interfaces impinge.

\begin{figure}
\begin{center}
\includegraphics[height=\textwidth, angle=-90]{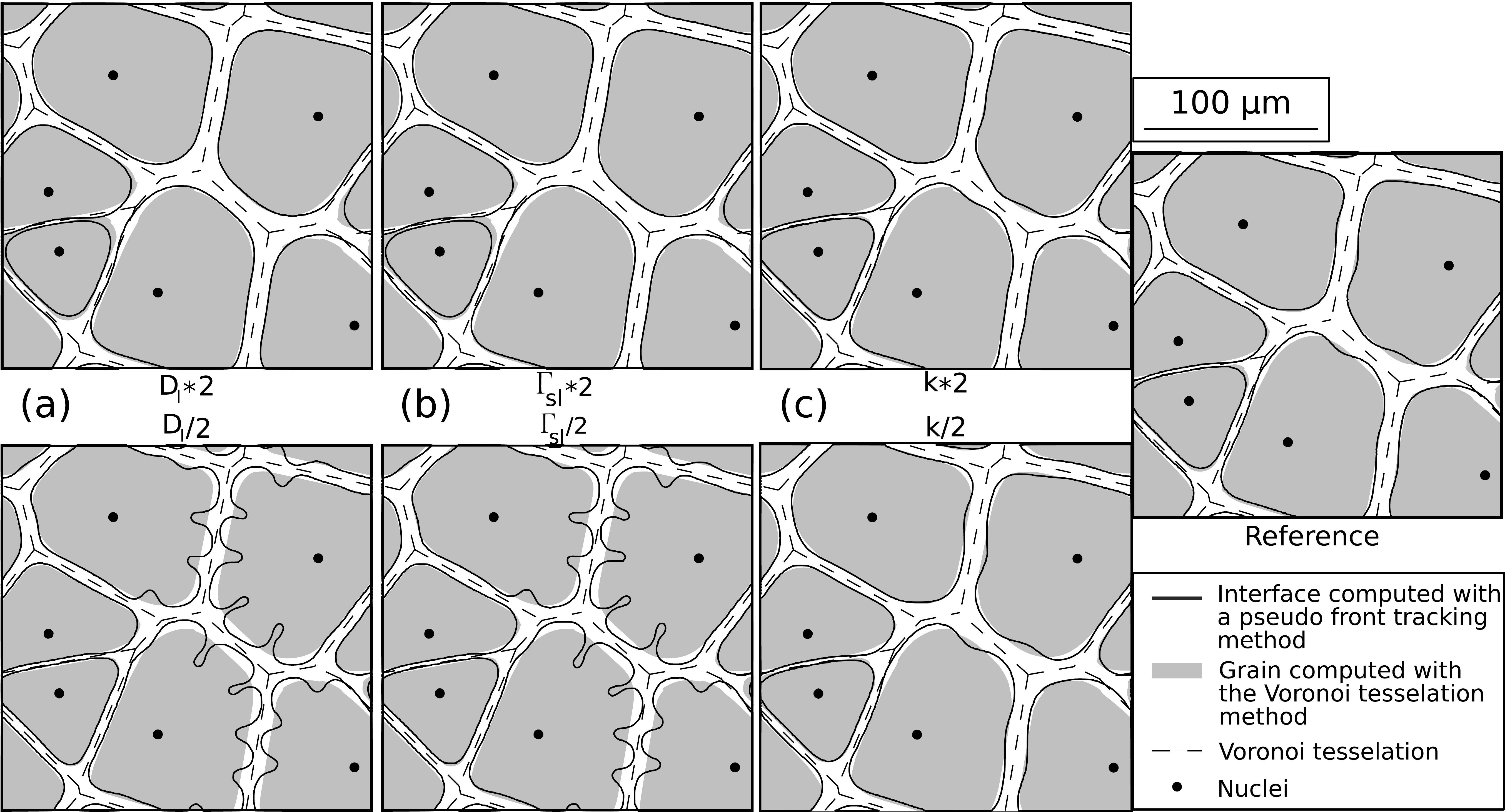}
\caption{Comparison of grain shape at $g_s$=0.68 for various sets
of solidification parameters. Starting from the reference
simulation (right figure), one parameter is changed at a time:
$D_\ell$ (a), $\Gamma_{s\ell}$ (b) and $k$ (c) are multiplied
(top) or divided (bottom) by 2 with respect to the reference.}
\label{param}
\end{center}
\end{figure}

Figure \ref{param} shows the effect of various parameters on the
shape of the solid liquid interfaces. It is convenient to
introduce a dimensionless number, $C$:

\begin{equation}
C=\frac{1}{d_c}\left( \frac{\Gamma_{s\ell}
D_\ell}{(-\dot{T})}\right) ^\frac{1}{3}
\end{equation}

This number is the ratio of a "typical" curvature radius at grains
corner (i.e., when $(\tan \alpha - \alpha)^{-1} = 0.5$) and the
average grain size. Figures \ref{param} a) and b) show that a
larger $C$ number corresponds indeed to a larger radius of
curvature at grains corners, regardless whether this is achieved
by increasing $D_\ell$ or $\Gamma_{s\ell}$. Moreover, the
agreement between the PFT and Voronoi predictions remains when the
physical parameters are changed. Yet, Fig. \ref{param} also points
out the limits of the present model, in particular at small $C$
number. Indeed, the PFT predicts a destabilization of the globular
grains into dendritic ones when $C$ decreases. A criterion for the
transition from globular to dendritic equiaxed grains has been
recently proposed by Diepers et al.\cite{trans_globu_dend} These
authors found that it occurs when:

\begin{equation}
d_c^{GD}=A_{GD} \left(\frac{L}{k \Delta T_o c_p} \frac{\Gamma_{s\ell}
D_\ell }{ (-\dot{T})}\right) ^\frac{1}{3}\label{Diepers}
\end{equation}

where $c_p$ and $L$ are the volumetric specific heat and latent of
fusion, respectively, and $A_{GD}$ a dimensionless factor 
function of the anisotropy of the solid-liquid interfacial energy.
Although this criterion has been derived for a sphere and a
cooling rate imposed only at the boundaries of the system, it is
interesting to note that the same power-law of the ratio ${D_\ell
\Gamma_{s\ell} \over \dot{T}}$ is retrieved. The additional
term,$(L/( k \Delta T_o c_p))$, comes from the solute undercooling ahead the solidification front and from an overall thermal balance. The destabilisation of the interface occurs at the beginning of solidification when this undercooling is crucial, whereas the selection of the grains corner curvature occurs latter in solidification when the solutal profile is flat.  A change
in the partition factor $k$, which does not affect the radius of
curvature of the grains calculated with Eq. \ref{R}, nevertheless
slightly influences the destabilization of the grains as predicted
by Eq. \ref{Diepers} from the factor $k\Delta T_o$ and as observed
with the PFT method (Fig. \ref{param} c).

For a high $C$ number, another limitation of the model is
encountered as illustrated by the solidification of the smallest
grain in Figs. \ref{param} a) b). The Voronoi model does not predict
well the solidification of this grain at high $C$ number. This is because the overall curvature of the grain is not accounted for in the Voronoi method.
\cor{Yet, the curvature undercooling during growth of globulitic grains remains very small, typically 0.01 K  for a spherical grain of 20 $\mu m$ radius. That of the last liquid droplets is slightly higher considering that the droplets are smaller and have a negative curvature, opposite to that considered here for the last liquid located at triple junctions. But in any case, curvature undercooling, despite its importance for coarsening and for the shape of the dendrite tip, is small in dendritic and globulitic solidification under normal conditions.} 

\cor{In order to use a Voronoi construction, e.g., linear grain boundaries, to predict the final grain structures, two assumptions have been made: first, the temperature is uniform at the scale of the grains, and second, the nuclei all start at the same time. The first hypothesis is usually verified for globulitic structures, e.g., grains of about 100 $\mu$m growing in a thermal gradient lower than 1 K/cm. The temperature difference across a reference specimen containing 100 grains is just 1 K in this case. The second hypothesis clearly} \cor{pertains to alloys which are inoculated. It has been shown in the case of Al-alloys inoculated with TiB$_2$ particles that nucleation is athermal \cite{athermal}. In athermal nucleation, the activation of a nucleant is a function of the undercooling only and not of time. In this case, the range of particle sizes which have been shown to be activated during solidification is typically 2-5 $\mu$m in diameter \cite{athermal}. Therefore, the maximum nucleation undercooling is about 0.2 K for a Gibbs-Thomson coefficient of $10^{-7}$ Km. This value is small compared to the growth undercooling and thus an instantaneous nucleation assumption appears reasonable.}

The most striking feature of the present Voronoi
model is its low computation cost. The PFT simulations presented
in Fig. \ref{interfaces} take around 12 Hours on a 2.8 GHz Pentium
4 personal computer, whereas the Voronoi calculations requires
less than 2 s! This represents a gain of more than 4 orders of
magnitude. This difference is even more striking with large mushy
zones. The computation of a whole mushy zone solidification that
contains 14000 grains (see Fig. 9 of Ref. \cite{second}) requires
less than 10 seconds on the same computer.

\section{Conclusion}
A model for the solidification of globular grains based on Voronoi
diagrams and round corners has been derived. This model is shown to estimate well the shape of the
globular grains during solidification and to take into account the
effect of the various physical parameters, providing the interface does not become unstable. Its very low
computation cost makes it an ideal base for granular simulation of
mushy zone at a mesoscale, that can compute macroscopic properties
such as mechanical properties or feeding properties based on the
behaviour of individual grains \cite{second}. Therefore, this model
can be easily extended to 3 dimensions, providing new opportunities for
realistic granular simulation.

\section{Acknowledgement}

This research is funded by Alcan CRV (France) and ANRT (Association Nationale de la Recherche
 Technique, France). The authors would like to
thank Dr. A. Jacot from LSMX EPFL (Switzerland) for his help with the PFT simulations.

\section{Appendix}
Considering a section of the flat interface far from the corner, the speed of the interface can be estimated from the following flux balance.
\begin{equation}
h\ppd{c_\ell^\infty}{t}= c_\ell^\infty (1-k) v^\infty -j_{bd}^{\infty}
\label{one}
\end{equation}
The left hand term represents the variation of solute in the liquid, where $h$ is the thickness of the liquid film. The first right hand term correspond to the solute rejected in the liquid due to the advance of the interface and  $j_{bd}^{\infty}$ is the flux pumped in the solid by back-diffusion. Similarly, a solute balance on the liquid part surrounding the corner (see main section) can be derived. If $\Omega$ denotes this domain, one has.

\begin{equation}
\int_\Omega \ppd{c_\ell}{t} ds= c_\ell^R (1-k) R \sin \alpha v^R - \alpha R j_{bd}^{R} + \Phi
\label{two}
\end{equation}
Please note that this balance accounts for the flux $\Phi$ exchanged between the flat and curved portions of the interface.

The variation of solute concentration at the interface is imposed by the cooling rate and 
$ \partial _t c_\ell^\infty=  \partial _t c_\ell^R =\dot{T}/{m}$. As a consequence, whatever is the 
precise repartition of solute around the grain corner, the variation of solute around the grain 
corner can be estimated by:
\begin{equation}
\int_\Omega \ppd{c_\ell}{t} ds= S_\Omega \frac{\dot{T}}{m}  
\label{three}
\end{equation}
where $S_\Omega$ is the area of the domain surrounding the round corner.
As $v^r= v^\infty / \cos \alpha$ (see main section), Eqs \ref{one}, \ref{two} and \ref{three} give:
\begin{equation}
S_\Omega \frac{\dot{T}}{m} = \frac{c_\ell^R} {c_\ell^\infty}  R \tan \alpha (h\frac{\dot{T}}{m}+j_{bd}^{\infty}) - \alpha R j_{bd}^{R} + \Phi
\end{equation}
Considering that $c_\ell^R / c_\ell^\infty \sim 1$ and neglecting the differences of back diffusion along the flat and curved parts of the interface, one gets:
\begin{equation}
(S+R h \tan \alpha) \frac{\dot{T}}{m} =  R h \tan \alpha \frac{\dot{T}}{m} + \Phi
\end{equation}
Where the surface $S_\Omega$ have been separated into the surface delimited by the extension of the flat interface $(Rh\tan \alpha)$ and an extra surface $S$ represented in grey in Fig. \ref{coin}. As stated in the main part, one retrieve the fact that the solute rejected by the flat interface moving at velocity $v^\infty$ is equivalent to that rejected by the curved interface moving at velocity $v^r= v^\infty / \cos \alpha$. Removing this term on the left and right hand sides finally gives: 
\begin{equation}
S \frac{\dot{T}}{m} =  \Phi
\end{equation}
This represents the solute balance between the solute flux induced by the Gibbs-Thomson effect and the geometrical advantage of a corner for diffusion.


\begin{thebibliography}{10}

\bibitem{Campbell1991}
J.~Campbell.
\newblock {\em Castings}.
\newblock Butterworth Heineman, 1991.

\bibitem{phase_field_rev}
W.~J. Boettinger, J.~A.Warren, C.~Beckermann, and A.~Karma.
\newblock Phase-field simulation of solidification.
\newblock {\em Annu. Rev. Matter. Res.}, 32:163--94, 2002.

\bibitem{jon_level_set}
Y.T. Kim, N.~Goldenfeld, and J.A. Dantzig.
\newblock {C}omputation of dendritic microstructures using a level set method.
\newblock {\em Phys. Rev. E}, 62(2):2471--74, 2000.

\bibitem{jacot}
A.~Jacot and M.~Rappaz.
\newblock A pseudo-front tracking technique for the modelling of solidification
  microstructures in multi-component alloys.
\newblock {\em Acta Mater.}, 50:1909--26, 2002.

\bibitem{percotheorie}
D.~Stauffer and A.~Aharony.
\newblock {\em Introduction to percolation theory}.
\newblock Taylor and Francis, 1994.

\bibitem{second}
S.~Vern\`ede, Ph. Jarry, and M.~Rappaz.
\newblock A granular model of equiaxed mushy zones: Formation of a coherent
  solid and localization of feeding.
\newblock {\em Acta Mater.}, 54:4023--34, 2006.

\bibitem{vince2}
V.~Mathier, A.~Jacot, and M.~Rappaz.
\newblock Coalescence of equiaxed grains during solidification.
\newblock {\em Mod. Sim. Mat. Sci. Eng.}, 12:479--490, 2004.

\bibitem{premier}
S.~Vern\`ede and M.~Rappaz.
\newblock Transition of the mushy zone from continuous liquid films to a
  coherent solid.
\newblock {\em Phil. Mag.}, 86(23):3779--94, 2006.

\bibitem{charbon}
Ch. Charbon and M.~Rappaz.
\newblock Shape of grain boundaries during phase transformations.
\newblock {\em Acta Mater.}, 44:2663--68, 1996.

\bibitem{qhull}
C.B. Barber, D.P. Dobkin, and H.T. Huhdanpaa.
\newblock The quickhull algorithm for convex hulls.
\newblock {\em ACM Trans. of Mathematical Software}, 22:469--83, 1996.

\bibitem{Landau_solidif}
V.R. Voller and S.~Sundarraj.
\newblock {\em Mater. Sci. Tech.}, 9:474--481, 1993.

\bibitem{coal}
M.~Rappaz, A.~Jacot, and W.~Boettinger.
\newblock Last stage solidification of alloys : Theorical model of dendrite arm
  and grain coalescence.
\newblock {\em Met. Mater. Trans.}, 34A:467--479, 2003.

\bibitem{organiques}
I.~Farup, J.M. Drezet, and M.~Rappaz.
\newblock In situ observation of hot tearing formation in
  succinonitrile-acetone.
\newblock {\em Acta Mater.}, 49:1261--69, 2001.

\bibitem{plic}
D.B. Kothe, W.J. Rider, S.J. Mosso, and J.S. Brock.
\newblock Technical report, Los Alamos Research Laboratories, 1996.

\bibitem{Carman}
P.C. Carman.
\newblock Fluid flow through granular beds.
\newblock {\em Trans. Inst. Chem.}, 15:150, 1935.

\bibitem{Ksv}
O.~Nielsen, L.~Arnberg, A.~Mo, and H.~Thevik.
\newblock Experimental determination of mushy zone permeability in
  aluminum-copper alloys with equiaxed microstructures.
\newblock {\em Met. Mater. Trans. A}, 30A:2455--62, 1999.

\bibitem{triple_point}
R.~Folch and M.~Plapp.
\newblock Quantitative phase-field modeling of two-phase solidification.
\newblock {\em Phys. Rev. E}, 72(1):011602, 2005.

\bibitem{trans_globu_dend}
H.~J. Diepers and A.~Karma.
\newblock {G}lobular-denditic transition in equiaxed alloy solidification.
\newblock In {\em Solidification Processes and Microstructures}. M. Rappaz,
  Ch.Beckermann, R. Trivedi, (TMS Publ., Warrendale, PA, USA), 2004.

\bibitem{athermal}
T.E. Quested and A.L. Greer.
\newblock Athermal heterogeneous nucleation of solidification.
\newblock {\em Acta Mater.}, 53:2683--2692, 2005.

\end{thebibliography}
\end{document}